\documentclass{article}
\usepackage[preprint]{neurips_2024}
\usepackage[utf8]{inputenc}
\usepackage[T1]{fontenc}
\usepackage{hyperref}
\usepackage{url}
\usepackage{booktabs}
\usepackage{amsfonts}
\usepackage{subfig}
\usepackage{nicefrac}
\usepackage{microtype}
\usepackage{xcolor}
\usepackage{amsmath}
\usepackage{natbib}
\usepackage{graphicx}

\title{AeTHERON: Autoregressive Topology-aware Heterogeneous Graph Operator Network for Fluid-Structure Interaction}

\author{%
  Sushrut Kumar\thanks{skumar94@jhu.edu} \\
  Department of Mechanical Engineering\\
  Johns Hopkins University\\
  Baltimore, MD 21218 \\
}

\begin{document}
\maketitle

\begin{abstract}
Surrogate modeling of body-driven fluid flows where immersed moving boundaries 
couple structural dynamics to chaotic, unsteady fluid phenomena remains a 
fundamental challenge for both computational physics and machine learning. We 
present AeTHERON, a heterogeneous graph neural operator whose architecture directly 
mirrors the structure of the sharp-interface immersed boundary method (IBM): a 
dual-graph representation separating fluid and structural domains, coupled through 
sparse cross-attention that reflects the compact support of IBM interpolation 
stencils. This physics-informed inductive bias enables AeTHERON to learn nonlinear 
fluid-structure coupling in a shared high-dimensional latent space, with continuous 
sinusoidal time embeddings providing temporal generalization across lead times. We 
evaluate AeTHERON on direct numerical simulations of a flapping flexible caudal 
fin — a canonical FSI benchmark featuring leading-edge vortex formation, large 
membrane deformation, and chaotic wake shedding across a $4 \times 5$ parameter 
grid of membrane thickness ($h^* \in \{0.01$--$0.04\}$) and Strouhal number 
($St \in \{0.30$--$0.50\}$). As a proof-of-concept, we train on the first 150 
timesteps of a representative case using a 70/30 train/validation split and 
evaluate on the fully unseen extrapolation window $t = 150$--$200$. AeTHERON 
captures large-scale vortex topology and wake structure with qualitative fidelity, 
achieving a mean extrapolation MAE of $0.168$ without retraining, with error 
peaking near flapping half-cycle transitions where flow reorganization is most 
rapid. \textit{This is a continuously
developing preprint; results and figures will be updated in subsequent versions.}
\end{abstract}

\section{Introduction}
Accurate and efficient simulation of body-driven flows such as in fluid-structure
interactions (FSI) remains one of the most challenging problems in computational physics
and engineering. Applications ranging from biological locomotion and bio-inspired
propulsion systems to autonomous underwater vehicles, flexible airfoils, and morphing
wing designs require high-fidelity models capable of capturing complex, unsteady, and
strongly nonlinear phenomena governed by coupled partial differential equations (PDEs)
and evolving boundary geometries. Traditional numerical approaches such as the immersed
boundary methods provides really high accuracy but often demand prohibitive computational
resources, especially for repeated simulation, design optimization, uncertainty
quantification, and real-time control applications
\cite{mittal2005immersed,griffith2012immersed,peskin2002immersed}.

The immersed boundary method (IBM), first introduced by Peskin
\cite{peskin2002immersed}, were developed to handle the moving body CFD simulations and
has been extended by numerous researchers
\cite{mittal2008versatile,mittal2025freeman,griffith2012immersed}. It has emerged as a
particularly powerful framework for simulating flows with complex, moving, and deforming
boundaries such as the flow within human heart \cite{viola2023gpu,viola2023high}, blood
vessels \cite{bailoor2021computational}, animal locomotion
\cite{kumar2025computational,seo2026scaling,zhou2024effect,zhou2025effect,zhou2025hydrodynamically},
incompressible \cite{menon2022contribution} and compressible aerodynamics
\cite{turner2024high}. These class of methods avoid the need for body-conforming grids
and enable robust handling of large deformations \cite{kumar2025computational} along
with multi-body interactions \cite{verzicco2021collision}. The computational cost of
high-fidelity IBM simulations remains substantial, particularly when exploring parameter
spaces or conducting design studies involving hundreds or thousands of geometric and
kinematic configurations — a bottleneck that is especially acute in bio-inspired
propulsion studies where the coupling between membrane flexibility and flow topology
must be resolved across a wide range of kinematic parameters
\cite{kumar2026flow,kumar2026gpu}.

Data-driven surrogate models have attracted significant attention to address the
computational expense of running high-fidelity calculation. Although, the solutions
aren't always guaranteed to be perfect \cite{taira2025machine,mcgreivy2024weak} but can
provide good estimates of fluid flow solutions for variety of engineering applications
\cite{nabian2025mixture,ranade2025domino,mousavi2025rigno}. Machine learning techniques
such as convolutional neural networks (CNNs) \cite{guo2016convolutional},
physics-informed neural networks (PINNs) \cite{raissi2019physics}, and most recently,
neural operators \cite{li2021fourier,lu2021learning,kovachki2023neural} and graph neural
networks (GNNs) \cite{franco2023deep,gao2024data}, are now being developed to
approximate the underlying solution manifold and dynamical evolution of physical systems.
Neural operator frameworks—including the Fourier Neural Operator (FNO)
\cite{li2021fourier}, Deep Operator Network (DeepONet) \cite{lu2021learning}, and
state-space neural operator architectures \cite{nguyen2024state}—have demonstrated
mesh-independent parameterization of solution maps in elasticity, fluid dynamics, and
phase-field evolution, offering the promise of resolution-invariant surrogate modeling.
Early work applying regression-based machine learning to extract physical quantities
from experimental fluid data \cite{basu2020quantitative} established the feasibility
of data-driven approaches for flow field reconstruction, motivating the more expressive
architectures explored here.

While global operator networks offer flexibility and rapid inference, recent studies
highlight their fundamental limitations in generalization and stability when training
data is scarce or when dynamics are governed by strongly local physical coupling. These
limitations are particularly pronounced in scenarios involving sharp gradients,
discontinuities, or multiscale phenomena—characteristics that are ubiquitous in
body-driven flows with boundary layer separation, vortex shedding, and wake interactions.
To address these deficits, graph neural simulators (GNS) and related GNN-based
surrogates have emerged as effective alternatives, leveraging locality and
message-passing inductive biases to mirror classical numerical stencils and support
robust, long-horizon rollouts
\cite{sanchez2020learning,pfaff2021learning,fortunato2022multiscale}. This approach has
shown particular promise in time-dependent systems where explicit time-stepping schemes
can provide superior stability compared to direct operator mappings \cite{nayak2025data}.

Despite this progress, surrogate modeling for body-movement-driven flows—with explicit
geometric coupling and immersed boundaries—remains relatively underexplored. Existing
GNN frameworks often focus on homogeneous domains with fixed topologies, limited emphasis
on mesh adaptation, or simplified treatment of interfacial coupling. More critically,
conventional approaches typically handle fluid-structure interactions through direct
feature concatenation or simple edge-based coupling, failing to capture the rich,
nonlinear interdependencies that characterize these multiphysics systems. The challenge
is compounded by the computational cost of generating sufficient high-fidelity training
data, which demands GPU-accelerated solvers capable of operating at extreme scales
\cite{kumar2026gpu}.

A key insight driving this work is that fluid-structure interactions in immersed boundary
systems are fundamentally mediated through complex, nonlocal relationships that are best
represented and learned in abstract latent spaces rather than through direct geometric
coupling. In biological systems, for instance, the motion of a fish's caudal fin drives
flow through intricate pressure-velocity coupling mechanisms that extend far beyond
immediate geometric neighborhoods \cite{gao2020predicting}. Passive membrane
deformation — where the fin's flexibility modulates the leading-edge vortex dynamics
and wake structure to enhance propulsive efficiency — creates a feedback loop between
structural compliance and fluid forcing that is both physically rich and computationally
expensive to resolve \cite{kumar2026flow}. Similarly, in engineering applications,
the deformation of flexible structures induces flow modifications that propagate through
multiple length scales and time scales, creating feedback loops that are difficult to
capture with purely local message-passing schemes.

This work presents AeTHERON, a novel heterogeneous graph neural operator that addresses
these challenges through an innovative latent space interaction paradigm. Our approach
recognizes that effective fluid-structure coupling requires learning abstract
representations of both domains that can interact through cross-attention mechanisms,
enabling the model to discover and exploit hidden correlations that govern the coupled
dynamics. Specifically, we propose that the motion of immersed boundaries drives fluid
evolution most naturally when both fluid and structural states are elevated to a shared
high-dimensional latent space where their interactions can be learned and modeled through
attention-based mechanisms
\cite{vaswani2017attention,wang2019heterogeneous,jaegle2021perceiver}.

Our test case involves a flapping membrane fin immersed in a viscous flow, spanning
regimes with strong transient behavior, nonlinear dynamics, and complex wake-shedding
phenomena. This configuration captures many of the challenging aspects of biological and
bio-inspired propulsion systems, including leading-edge vortex formation, boundary layer
separation, and unsteady force generation mechanisms. The high-fidelity training data is
generated using a GPU-accelerated sharp-interface IBM solver \cite{kumar2026gpu} capable
of resolving these dynamics at the scale required for surrogate training.

The methodological advances and computational tools developed herein are readily
extensible to broader classes of multiphysics problems and offer a foundation for
next-generation AI-accelerated simulation, design, and control with physics-driven
constraints.cs-driven
constraints.

\section{Problem Formulation and Model Description}

We address the challenge of developing a neural operator to estimate the fluid flow
solution governed by the incompressible Navier-Stokes equation by combining graph-based
neural networks with immersed boundary methods (IBMs). Operator learning requires that
the underlying network structure be agnostic to the computational grid and that the
inference transferability be achieved with a single trained instance of the model. We
achieve this task by representing the problem using two heterogeneous graphs akin to
immersed boundary methods, implemented using the heterogeneous graph learning suite
within the PyTorch Geometric library.

Here, we make use of two graph structures: the fluid graph
$\mathcal{G}_f = (\mathcal{V}_f, \mathcal{E}_{f \to f})$ and the membrane graph
$\mathcal{G}_m = (\mathcal{V}_m, \mathcal{E}_{m \to m})$. $\mathcal{V}_f$ is the set
of fluid nodes with features $\mathbf{x}^f_i(t) \in \mathbb{R}^{d_f}$ (e.g., velocity),
and $\mathcal{E}_{f \to f}$ contains edges $(i, j)$ with attributes
$\mathbf{e}_{i,j}^f \in \mathbb{R}^{d_{e,f}}$. Similarly, $\mathcal{V}_m$ is the set
of membrane nodes with features $\mathbf{x}^m_p(t) \in \mathbb{R}^{d_m}$ (e.g.,
displacement), and $\mathcal{E}_{m \to m}$ contains edges $(p, q)$ with attributes
$\mathbf{e}_{p,q}^m \in \mathbb{R}^{d_{e,m}}$. Interaction between the two domains is
modeled through cross-graph edges
$\mathcal{E}_{m \to f} \subseteq \mathcal{V}_m \times \mathcal{V}_f$, with edges
$(p, i)$ and attributes $\mathbf{e}_{p,i}^{m \to f} \in \mathbb{R}^{d_{e,mf}}$. This
iteration focuses on modeling the expensive CFD component; the structural solve is
orders of magnitude cheaper and the FSI-generated membrane shapes are used to prescribe
the body motion for the surrogate.

The goal is to obtain a finite-dimensional approximation of the neural operator
$\mathcal{N}_\theta$ to obtain the evolved fluid flow state at lead time $\tau$:
\begin{equation}
    \mathcal{N}_\theta: \mathcal{F}_{in}(t) \rightarrow \mathcal{F}_{out}(t+\tau)
    \label{eq:NO}
\end{equation}
\begin{equation}
    \mathbf{x}^f(t+\tau) = \mathcal{N}_\theta(\mathbf{x}^f(t), \mathbf{x}^m(t), \tau)
    = \Psi_f \circ(\mathcal{M}_l\circ \cdots \circ \mathcal{M}_1) \circ
    \Phi_f(\mathbf{x}(t),\tau)
    \label{ref:NNeq}
\end{equation}

\begin{figure}[ht]
    \centering
    \includegraphics[width=1\textwidth]{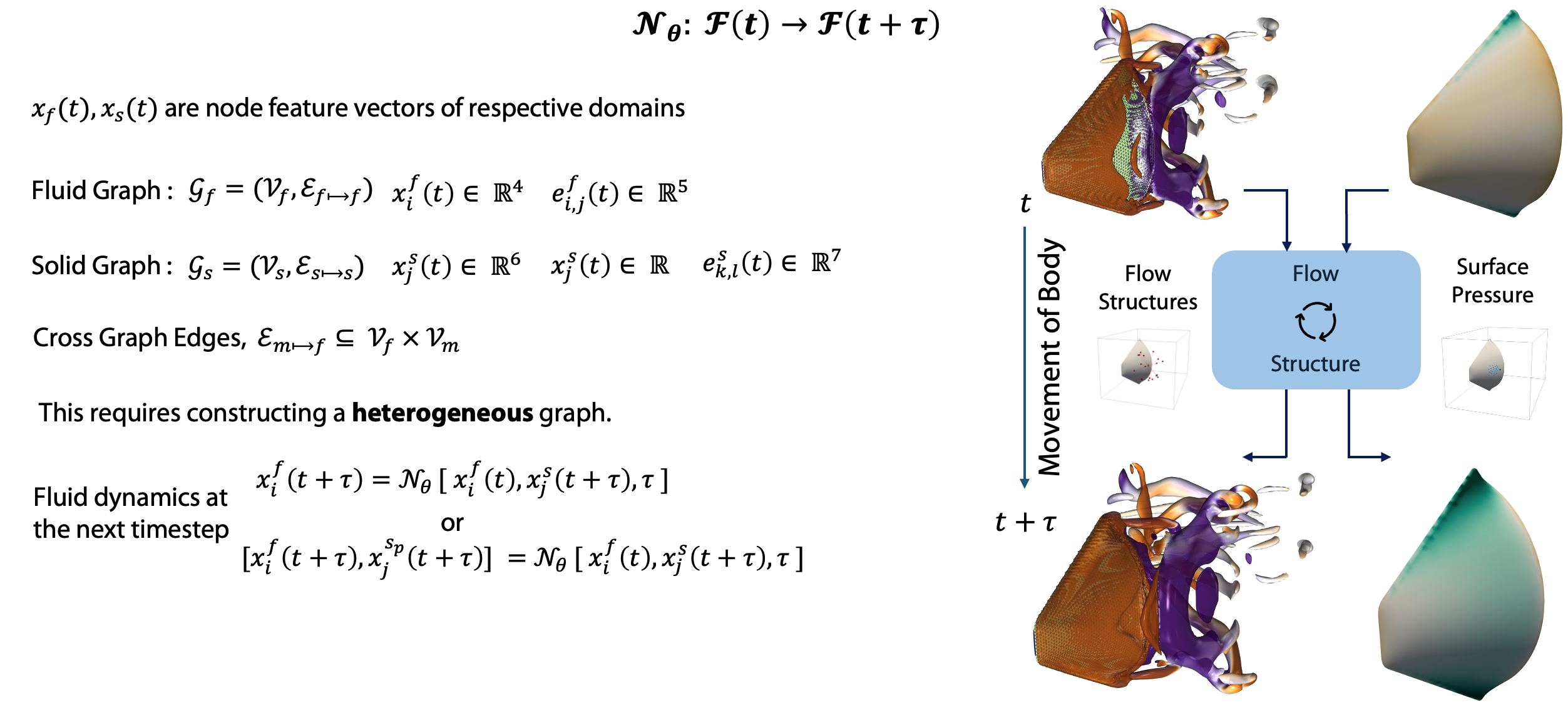}
    \caption{AeTHERON problem formulation and physical motivation. \textbf{Left:} The
    FSI surrogate is cast as a neural operator
    $\mathcal{N}_\theta: \mathcal{F}(t) \rightarrow \mathcal{F}(t+\tau)$ operating on
    a heterogeneous graph pair — a fluid graph
    $\mathcal{G}_f = (\mathcal{V}_f, \mathcal{E}_{f \to f})$ with 4-dimensional node
    features and 14-dimensional edge attributes, and a solid graph
    $\mathcal{G}_s = (\mathcal{V}_s, \mathcal{E}_{s \to s})$ with 10-dimensional node
    features and 7-dimensional edge attributes — coupled through cross-graph edges
    $\mathcal{E}_{m \to f} \subseteq \mathcal{V}_f \times \mathcal{V}_m$. Given current
    fluid state $\mathbf{x}^f_i(t)$ and membrane state at lead time
    $\mathbf{x}^s_j(t+\tau)$, the operator predicts the evolved fluid state
    $\mathbf{x}^f_i(t+\tau)$. \textbf{Right:} Physical motivation for the heterogeneous
    graph design. The body motion drives coupled evolution of flow structures (vortex
    isosurfaces, colored by spanwise vorticity) and surface pressure distribution at
    time $t$ (top) and $t+\tau$ (bottom). The bidirectional flow-structure feedback is
    encoded in AeTHERON through the cross-graph sparse attention mechanism, mirroring
    the sharp-interface IBM interpolation stencil at the numerical level.}
    \label{fig:architecture}
\end{figure}

The operator uses three components: encoder ($\Phi$), processor ($\mathcal{M}_l$), and
decoder ($\Psi$), illustrated in Figure~\ref{fig:architecture}.

\subsection{Encoder}
The input features of the membrane and the flow can lie in vector spaces of different dimensions. Therefore, we design an encoder stage to project the input features from the physical domains into a shared high-dimensional latent space. This enables us to model the complex nonlinear interactions between fluid flow and membrane motion within a shared space. Moreover, this lifting enriches the input feature vector in the high-dimensional space, enabling the model to learn interactions more effectively. Therefore, with this approach, the latent representation of the membrane feature drives the latent fluid-flow feature, consistent with our physical intuition that the motion of the immersed body drives the evolution of the flow in the physical domain.

An important point is that the dataset's features lack the temporal information required for time forecasting. Incorporating this information is critical, as time-scales vary across datasets due to different flapping frequencies. This work makes use of the sinusoidal time embeddings, $\mathbf{e}(\tau) \in \mathbb{R}^{d_t}$ and concatenated to the physical features of both membrane and fluid flow. It is shown by $[ \cdot, \cdot ]$ in equation \ref{eq:timeEmbed}.
\begin{align}
    e (\tau) = \left[ \sin\left(2\pi \tau / 10^{k/(d_t/2)}\right), \cos\left(2\pi \tau / 10^{k/(d_t/2)}\right) \right]_{k=0}^{d_t/2-1}.
    \label{eq:timeEmbed}
\end{align}
This formulation is inspired by the EGNO work \cite{xu2024equivariant} and Fourier features to provide a continuous and bounded encoding of the lead time $\tau$.

For a membrane node \(p \in \mathcal{V}_m\) with features \(\mathbf{x}^m_p(t) \in \mathbb{R}^{d_m}\) and lead-time $\tau \in \mathbb{R}$, the encoder computes
\begin{equation}
    \mathbf{\xi}^m_p = \Phi_m \left( [\mathbf{x}^m_p(t+\tau), \mathbf{e}(\tau)] \right), \quad \Phi_m: \mathbb{R}^{d_m + d_t} \to \mathbb{R}^{d_h},
    \label{eq:membLatent}
\end{equation}

Here $\Phi_f$ is a feedforward neural network consisting of fully connected layers with SiLU activations \(\sigma(x) = x / (1 + e^{x})\) such that $\xi_m \in \mathbb{R}^{d_h}$. $[\mathbf{x}^m_p(t+\tau), \mathbf{e}(\tau)]$ is a concatenation of membrane features at lead time with time embeddings. We use the membrane feature at lead-time to inform the model of the membrane deformation state and let it learn the evolution of flow features. This is analogous to explicit time-stepping in numerical solvers.

For fluid nodes \(i \in \mathcal{V}_f\), the input features \(\mathbf{x}_v \in \mathbb{R}^{d_f}\) and time \(\tau \in \mathbb{R}\) are processed as:
\begin{equation}
    \mathbf{\xi}^f_i = \Phi_f \left( [\mathbf{x}^f_i(t), \mathbf{e}(\tau)] \right), \quad \Phi_f: \mathbb{R}^{d_f + d_t} \to \mathbb{R}^{d_h}.
\end{equation}

where $\Phi_f$ is a feedforward neural network such that $\xi_f \in \mathbb{R}^{d_h}$. By projecting both domains into a shared latent space of dimension \(d_h \gg d_f, d_m\), the encoders facilitate the alignment of fluid and membrane representations, modeling the physical coupling where membrane motion drives fluid evolution.

\subsection{Processor}
Next, we define the processor which comprises a stack of $L$ evolution layers. It is used to model the evolution of fluid flow ($\xi ^f_i$) under the influence of the prescribed membrane features ($\xi ^s_i$). Each evolution layer draws inspiration from Graph Neural Operator framework by Anandkumar et al. \cite{anandkumar2020neural}, where the model is based upon the iterative update as defined by equation \ref{eq:iterativeUpdate}.
\begin{align}
   \xi^{l+1}(x)=\sigma\left(W\xi^l(x)+\int_{D}\kappa\big(x,y,a(x),a(y)\big)\xi^l(y)dy\right), 
   \label{eq:iterativeUpdate}
\end{align}
Equation \ref{eq:iterativeUpdate} can be written using the Monte Carlo sum with a message passing neural network and has been successfully applied in various studies \cite{xu2025taylor}, including MAGNO, GKN, etc. In practice, the integral is estimated using a Monte Carlo approximation. Moreover, the presence of an immersed body should be taken into account. This is done by decomposing the discrete approximation into two aggregation operations, as shown in equation \ref{eq:monteCarlo}.
\begin{align}
   \xi^{f,l+1}_i =  \sigma\left(W_n\xi^{f,l}_i + A_i + A_c\right)
    \label{eq:monteCarlo}
\end{align}
where \(\sigma\) is SiLU, and \(W_l\) is a learnable matrix capturing local linear transformations. The two message aggregation operations are referred as intra-message passing ($A_i$) and cross-message passing ($A_c$). These operations will then involve neighborhood sets of solid and fluid points, $N_s(i)$ and $N_f(i)$ respectively, around the fluid point $i$ where $|\cdot|$ is the cardinality of the set. The operation for intra-message passing within the fluid domain models self-interactions akin to diffusion or advection terms in the Navier-Stokes PDE. The implementation for intra-message passing was adapted from Anandkumar et al.\cite{anandkumar2020neural} and is written as
\begin{equation}
    A_i = \frac{1}{|N_f(i)|} \sum_{j \in N_f(i)} \kappa_{f \to f} (\mathbf{e}^f_{i,j}) \mathbf{\xi}^{f,l}_j,
\end{equation}
where \(\kappa_{f \to f}: \mathbb{R}^{d_{e,f}} \to \mathbb{R}^{d_h \times d_h}\) is a parameterized kernel function that learns a matrix-valued kernel from edge attributes, $\mathbf{e}^f_{i,j}$. Next, the cross-message passing captures the influence of the prescribed membrane motion on the fluid flow (e.g., no-slip conditions), can be written as
\begin{equation}
    A_c = \frac{1}{|N_s(i)|}\sum_{k\in N_s}\kappa_{s\to f}\big(\xi^{f,l}_i,\xi^{s,l}_k\big)\xi^{s,l}_k,
\end{equation}
where $\kappa_{s\to f}$ is kernel function to model the interaction between flow features and solid features. The combined update can then be written as
\begin{align}
   \xi^{f,l+1}_i =  \sigma\Bigg[W_n\xi^{f,l}_i + 
    \underbrace{\frac{1}{|N_f (i)|}\sum_{j\in N_f}\kappa_{f\to f}\big(e^f_{i,j}\big)\xi^{f,l}_j}_{\text{Inter message passing}, A_i} +
    \underbrace{\frac{1}{|N_s(i)|}\sum_{k\in N_s}\kappa_{s\to f}\big(\xi^{f,l}_i,\xi^{s,l}_k\big)\xi^{s,l}_k}_{\text{Cross message passing}, A_c}
    \Bigg]
    \label{eq:monteCarlo}
\end{align}

To enhance cross-domain interaction, we parameterize the cross-message passing by drawing inspiration from the attention mechanism proposed by Vaswani et al. \cite{vaswani2017attention}. Specifically, we employ sparse attention due to the localized interactions. This sparsity, enforced by neighborhood \(N_m(i)\), reduces computational complexity from \(O(|\mathcal{V}_f| \cdot |\mathcal{V}_m|)\) to \(O(|\mathcal{E}_{m \to f}|)\), where \(|\mathcal{E}_{m \to f}|\) is typically much smaller due to local connectivity. This mirrors the compact support of regularized delta functions in diffuse IBMs \cite{peskin2002immersed} or interpolation stencils in sharp IBMs \cite{mittal2008versatile}, ensuring that only proximate membrane nodes influence each fluid node. This is done numerical by reinterpreting the $A_c$ using equation \ref{eq:attention-cross}
\begin{equation}
A_c = \frac{1}{|N_s|}\sum_{k \in N_s}\alpha_{i,k} V 
\label{eq:attention-cross}
\end{equation}
Here, $\alpha_{i,k} \in \mathbb{R}$ is the attention score. The attention scores allow learning the pairwise influences between fluid queries and solid keys. It then assists in weighted message passing using the value vector $V$, which is the attention-modulated version of solid features $\xi^{s,l}k$. For each fluid point $i$ and solid neighbor $k \in N_s(i)$, we define query $Q_i$, key $K_k$, and value $V_k$ as
\begin{align}
Q_i &= W_Q(\xi^{f,l}i), \quad K_k = W_K(\xi^{s,l}k), \quad V_k = W_V(\xi^{s,l}k),
\label{eq:attention-proj}
\end{align}

Here $W_Q, W_K, W_V: \mathbb{R}^{d_h} \to \mathbb{R}^{d_A}$ are feedforward neural networks that project features into the attention space of dimension $d_A$. The scaled similarity between $Q_i$ and each $K_k$ is
\begin{align}
    S_{i,k} = \frac{Q_i^T K_j}{\sqrt{d_A}}, S \in \mathbb{R}
\end{align}
The similarity is converted to score $\alpha_{i,k}$ using a softmax function, $\alpha_{i,k} = \text{softmax}(S_{i,k})$. 
Hence, the final update to latent fluid feature can be written using equation \ref{eq:flowLatentEvolve}.
\begin{align}
   \xi^{f,l+1}_i =  \sigma\Bigg[W_n\xi^{f,l}_i + 
    \underbrace{\frac{1}{|(N_f(i)|}\sum_{j \in N_f}\kappa_{f\to f}\big(e^f_{i,j}\big)\xi^{f,l}_j}_{\text{Inter message passing}, A_i} +
    \underbrace{\frac{1}{|N_s(i)|}\sum_{k \in N_s}\alpha_{i,k} V_k}_{\text{Cross message passing}, A_c}
    \Bigg]
    \label{eq:flowLatentEvolve}
\end{align}
After the message aggregation, a time-conditioning is performed by scaling and shifting the latent flow feature vector using a parameterized representation of the lead time \ref{eq:timeconditioning}.

\begin{align}
\xi^{f,t+1}_i = (1 + \tau \gamma(\tau))\odot\text{LayerNorm}(\xi^{f,l+1}_i) + \tau \lambda(\tau)
\label{eq:timeconditioning}
\end{align}

where $\gamma(\tau), \lambda(\tau): \mathbb{R} \to \mathbb{R}^{d_h}$ are neural networks with SiLU activations.

\subsection{Decoder}
The decoder stage reconstructs the physical fluid state from the final latent representation. The decoder design can be used either to obtain the fluid flow state at lead time or to get an incremental update, similar to a time-marching scheme, such as the Euler method. We utilize a time-marching update in our autoregressive predictions. The decoder update can be written as:
\begin{align}
    \mathbf{x}_i^{f,t+1} = \mathbf{x}_i^{f,t} + \tau \psi_f(\xi^{f,t+1}_i)
\end{align}
where \(\psi_f: \mathbb{R}^{d_h} \to \mathbb{R}^{d_f}\) is a feedforward neural network.

\subsection{Training Data and Procedure}

There are numerous examples of fluid-structure interaction, and we have chosen a thin
membrane-based caudal fin model flapping in a flow with an incoming velocity vector as
an application problem for this work. The problem has several complexities, such as
large membrane deformation, chaotic flow structures interacting with the body, and
varying flow physics with flapping frequency. The details regarding the caudal fin CFD
model are presented in Appendix~\ref{app:caudalFinModel}.

The full training dataset spans 20 direct numerical simulations covering a $4 \times 5$
parameter grid: four membrane thickness values ($h^* \in \{0.01, 0.02, 0.03, 0.04\}$)
and five Strouhal numbers ($St \in \{0.300, 0.325, 0.350, 0.450, 0.500\}$), covering
both low-frequency and high-frequency flapping regimes. Each simulation provides 200
timesteps of quasi-periodic coupled fluid-membrane data sampled at case-specific
intervals after the flow reaches statistical stationarity.

For the proof-of-concept results reported here, we train on a single representative
case ($h^* = 0.02$, $St = 0.40$). The 200 timesteps are split into a training window
($t = 0$--$149$, using a 70/30 random train/validation split) and a held-out
extrapolation window ($t = 150$--$200$) that is never seen during training. Results
across the complete $4 \times 5$ parameter grid will be reported in a subsequent
version.

Each simulation is represented as a heterogeneous graph pair: a fluid graph with
$|\mathcal{V}_f|$ nodes carrying 4-dimensional velocity features, and a membrane graph
with $|\mathcal{V}_m| = 1{,}757$ nodes carrying 10-dimensional displacement and
kinematic features. Graph edges are constructed using a neighborhood radius of $r = 0.04$
for all three edge types, consistent with the sharp-interface IBM interpolation stencil
width. The 14-dimensional fluid edge attributes encode relative positions, distances,
and local geometric quantities; the 7-dimensional membrane edge attributes encode
structural connectivity and stiffness information.

Training is performed using the Adam optimizer with an initial learning rate of
$2 \times 10^{-3}$, decayed by a factor of $0.5$ every 100 epochs over 200 epochs
total. The loss function is a weighted combination of MAE on the fluid velocity field
(weight 0.6) and membrane displacement field (weight 0.4). Batches of size 2 consist of
$(t, \tau)$ pairs drawn uniformly from the training cycles. All training was performed
on NVIDIA A100 and L40s GPUs at the ARCH HPC facility, Johns Hopkins University
(Rockfish cluster).

\section{Results}

\subsection{Learning Dynamics}


\begin{figure}[ht]
    \centering
    \includegraphics[width=\textwidth]{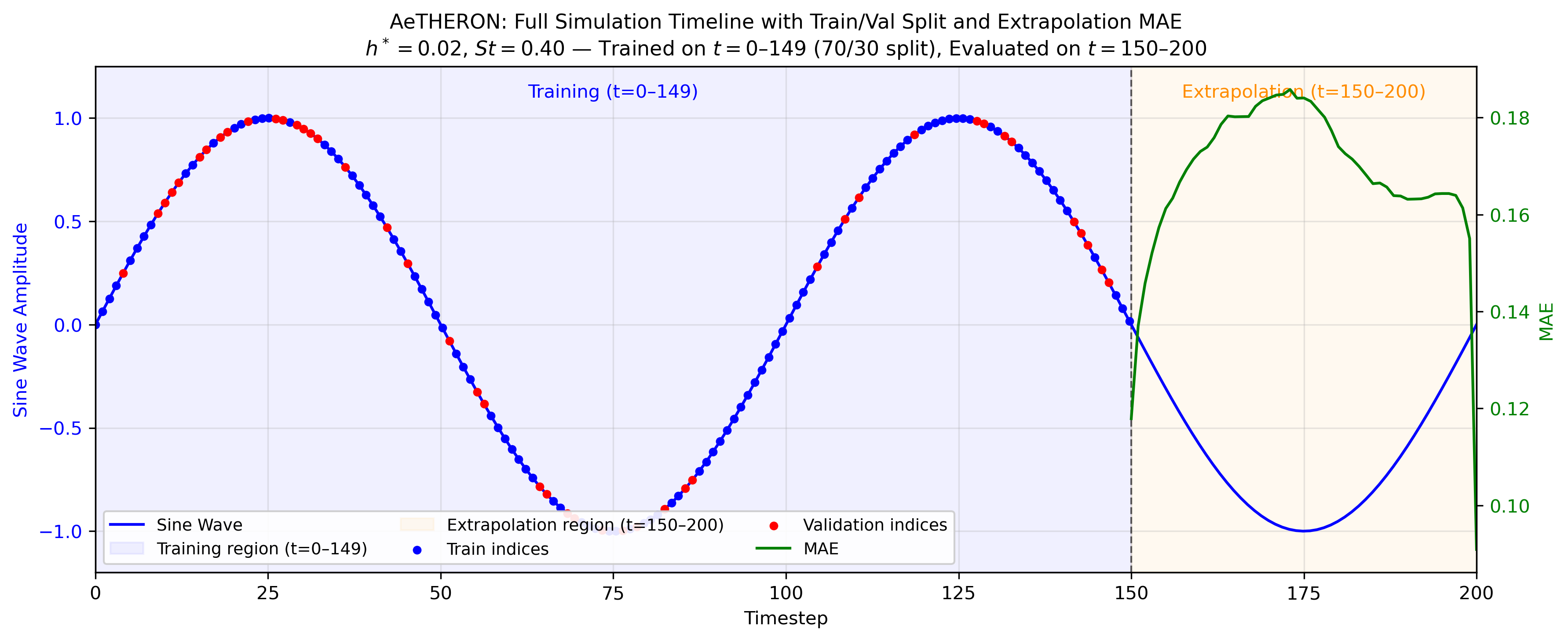}
    \caption{Full simulation timeline showing the train/validation split and extrapolation
    performance of AeTHERON ($h^* = 0.02$, $St = 0.40$). The sine wave (blue) represents
    the prescribed flapping cycle over 200 timesteps covering two full periods. Blue and
    red markers indicate training ($\approx$70\%) and validation ($\approx$30\%) indices
    within the training window $t = 0$--$149$. The model is then evaluated on the fully
    unseen extrapolation window $t = 150$--$200$ (orange region), where the MAE (green,
    right axis) ranges from $0.091$ to $0.186$ with a mean of $0.168$.}
    \label{fig:mae_timeline}
\end{figure}

We evaluate AeTHERON on a single representative caudal fin FSI case ($h^* = 0.02$,
$St = 0.40$) as a proof-of-concept demonstration. The full dataset spanning the
$4 \times 5$ parameter grid described in Appendix~\ref{app:dataset} is available for
training; results across the complete parameter space will be reported in a subsequent
version of this work.

The training window ($t = 0$--$149$) covers two flapping cycles split 70/30 into
training and validation sets using random sampling (Figure~\ref{fig:mae_timeline}). The
first flapping cycle ($t = 0$--$\approx 100$) captures the transient startup phase as
the flow develops from rest; the second cycle ($t \approx 100$--$149$) provides
fully-developed, quasi-periodic flow states. After training, the model is evaluated on
the held-out extrapolation window $t = 150$--$200$, which was never seen during training.

The MAE curve in Figure~\ref{fig:mae_timeline} reveals physically interpretable behavior.
The error rises from $0.118$ at $t = 150$ as the model begins predicting beyond its
training horizon, peaks at $0.186$ near $t = 170$ — corresponding to the flapping
half-cycle transition where leading-edge vortex formation and wake shedding drive rapid
topological changes — and then decreases to a minimum of $0.091$ at $t = 200$ as the
flow returns to a more quiescent phase. This phase-correlated error pattern is consistent
with the physical intuition that AeTHERON faces its greatest challenge when coupled
fluid-membrane dynamics are most strongly nonlinear. The mean extrapolation MAE of
$0.168$ is achieved without any retraining or fine-tuning on the prediction window.

\subsection{Prediction Results}

\begin{figure}[ht]
    \centering
    \includegraphics[width=\textwidth]{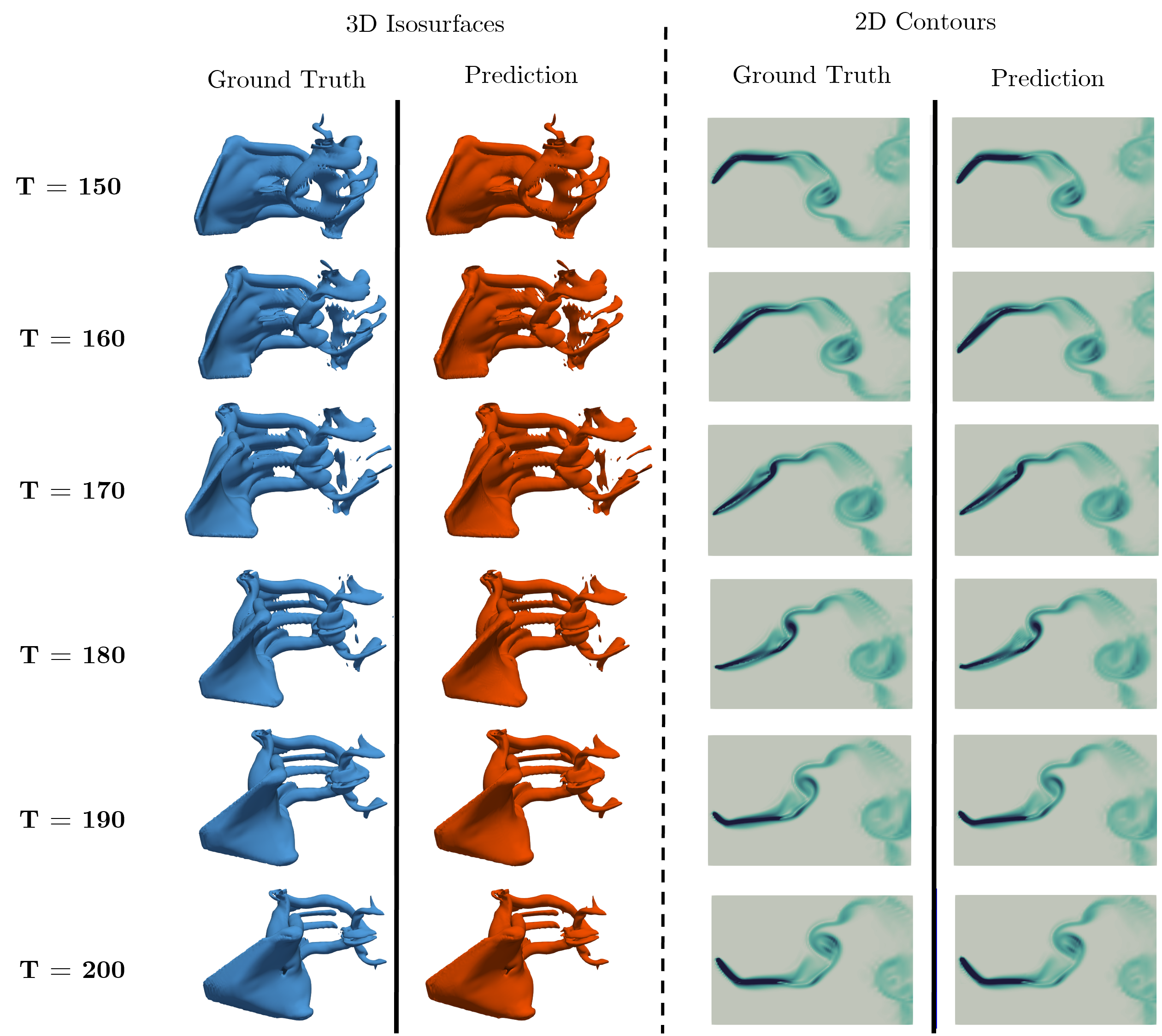}
    \caption{Qualitative comparison of AeTHERON predictions (orange) against ground
    truth DNS (blue) at six representative timesteps in the extrapolation window
    ($t = 150$--$200$). \textbf{Left two columns:} 3D isosurface visualizations of the
    vorticity magnitude showing large-scale vortex structures and wake topology.
    \textbf{Right two columns:} 2D cross-sectional slices of the vorticity magnitude in the
    mid-span plane; ground truth (left of each pair) vs.\ AeTHERON prediction (right).
    Rows progress from early ($t \approx 150$) to late ($t \approx 200$) in the
    extrapolation window, capturing multiple phases of the flapping cycle.}
    \label{fig:flow_comparison}
\end{figure}

Qualitative prediction results are presented in Figure~\ref{fig:flow_comparison},
showing 3D isosurface visualizations and 2D cross-sectional vorticity slices at six
representative timesteps across the extrapolation window. AeTHERON successfully captures
the large-scale vortex topology throughout the extrapolation window. The characteristic
arch and horseshoe vortex structures that dominate the near-wake of the flapping fin are
reproduced with qualitative fidelity across all six timesteps. The 2D cross-sectional
slices demonstrate particularly strong agreement — the predicted vorticity field closely
matches the ground truth in both the shape of the leading-edge vortex and the downstream
wake structure, including the roll-up and shedding of vortex pairs.

The most visible discrepancies appear in the fragmentation of fine-scale vortex filaments
at the tips and trailing edges, particularly in the upper rows where the flow undergoes
rapid reorganization. These regions correspond to the high-MAE phase near $t = 170$ in
Figure~\ref{fig:mae_timeline}, confirming that the primary limitation of AeTHERON lies
in resolving sharp, localized flow features during topological transitions rather than
in capturing global flow structure. The 2D slices show that even in these challenging
phases, the predicted large-scale vorticity patterns remain physically plausible and
structurally coherent.

Despite operating entirely outside its training distribution, AeTHERON maintains
qualitative fidelity across the full extrapolation window. Each forward pass completes
in milliseconds on GPUs, compared to hours of wall-clock time for the equivalent
DNS computation, demonstrating the practical utility of the surrogate for rapid flow
field estimation.

\section{Conclusion}

We presented AeTHERON, a heterogeneous graph neural operator for surrogate modeling of
body-driven fluid flows governed by the immersed boundary method. The key contribution
is an architecture whose graph topology and message-passing structure directly mirror the
numerical discretization of the IBM — a dual-graph representation separating fluid and
membrane domains, connected through sparse cross-attention that reflects the compact
support of sharp-interface interpolation stencils. This physics-informed architectural
inductive bias distinguishes AeTHERON from generic GNN surrogates that apply standard
message passing to FSI problems without accounting for the underlying numerical structure.

The sparse cross-attention mechanism reduces the fluid-structure coupling complexity from
$\mathcal{O}(|\mathcal{V}_f| \cdot |\mathcal{V}_m|)$ to
$\mathcal{O}(|\mathcal{E}_{m \to f}|)$, enabling scalable inference on realistic IBM
grids. The continuous sinusoidal time embedding enables temporal generalization. A
proof-of-concept evaluation on the caudal fin FSI benchmark demonstrates qualitative
fidelity in capturing vortex topology and wake dynamics across an unseen extrapolation
window, with mean extrapolation MAE of $0.168$ and inference speeds orders of magnitude
faster than the DNS solver.

\textbf{Limitations and future work.} The primary limitation is rollout stability during
topological vortex events. Future work will address this through multiscale message
passing for sharp-gradient regions near the immersed boundary, and physics-constrained
loss terms enforcing no-slip and no-penetration conditions explicitly. Extension to fully
coupled FSI — jointly predicting membrane deformation and flow field — is a natural next
step, as is application to cardiovascular FSI including heart valve dynamics and aortic
hemodynamics for patient-specific surgical planning.

\textbf{Broader impact.} AeTHERON provides a foundation for AI-accelerated simulation
in bio-inspired engineering, cardiovascular medicine, and fluid-structure interaction
design. Fast surrogate models for IBM-based FSI have direct applications in surgical
planning, device optimization, and real-time digital twin systems.

\begin{ack}
The author thanks Prof.\ Rajat Mittal and Dr.\ Jung-Hee Seo (Johns Hopkins University,
Flow Physics and Computation Laboratory) for guidance, the FSI simulation framework,
and the high-fidelity training data generated as part of ongoing research in the
laboratory. Computational resources were provided by the ARCH High Performance Computing
facility at Johns Hopkins University (Rockfish cluster, GPU partition). The author
acknowledges Josh Romero and Massimiliano Fatica (NVIDIA HPC) for collaboration on
GPU-accelerated simulation methods that informed the data generation pipeline for this
work. The author is grateful to Dr.\ Kamyar Azizzadenesheli for early discussions and to
Dr.\ Charlelie Laurent (NVIDIA PhysicsNeMo) for valuable discussions on autoregressive
rollout stabilization strategies that shaped the decoder design. 
\end{ack}

\bibliographystyle{plainnat}
\bibliography{references}


\appendix

\section{Details of Numerical Simulations for Data Generation}
\label{app:caudalFinModel}

\subsection{Caudal Fin Model}
Caudal fin refers to the propulsion elements used by fishes, where the organism flaps
the fin through undulation of the body. This undulating motion generates vortex
structures that induce pressure force vectors that provide thrust to the fish. The flow
physics involved in this case find many similarities with problems such as propellers,
heart valves, parachutes, and many more. This makes it a perfect case for applying the
deep learning model and building a surrogate.

In the current model, we assume the fin shape is a trapezoid with rigid leading, dorsal,
and ventral margins. The prescribed motion at these rigid regions of the fin drives the
entire fin. The fin then interacts with the incoming flow and deforms, which is governed
by a membrane deformation model. This work focuses on surrogate modeling of the expensive
fluid flow simulation; the structural solve is orders of magnitude faster and the
FSI-generated membrane shapes prescribe the body motion.

\begin{figure}[ht]
    \centering
    \subfloat[]{\includegraphics[width=0.45\textwidth]{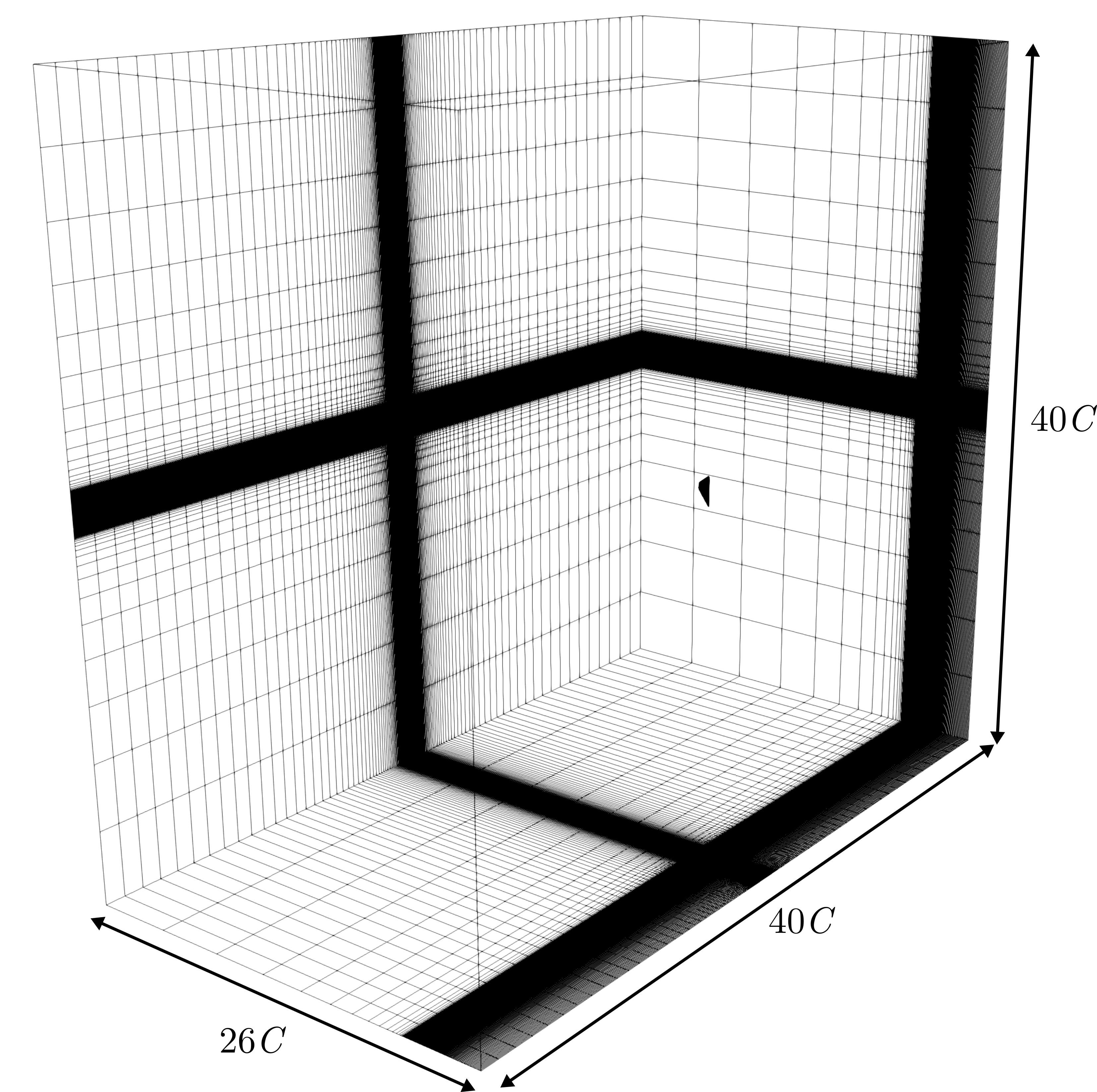}}
    \subfloat[]{\includegraphics[width=0.45\textwidth]{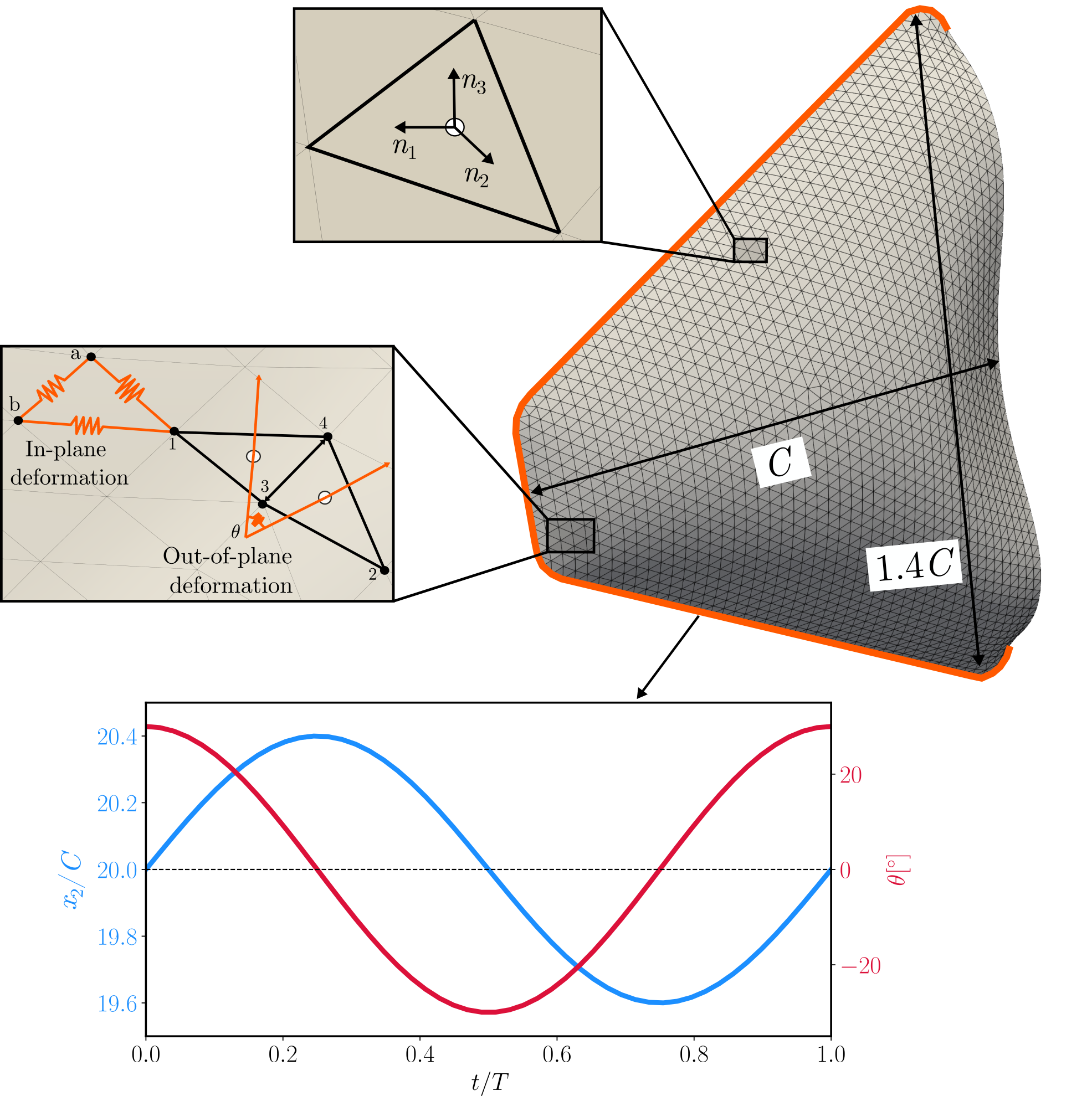}}
    \caption{Schematics of computational setup. (a) Rectilinear mesh of the flow domain,
    (b) Unstructured mesh of the membrane surface along with leading-edge kinematics,
    placement of spring network, and alignment of cell centroid normals.}
    \label{fig:compSetup}
\end{figure}

\subsection{Numerical Methods}
High-quality data is generated by direct numerical simulation of the incompressible
fluid flow governed by the Navier-Stokes equations:
\begin{align}
    \frac{\partial u_i}{\partial x_i} = 0; \quad
    \frac{\partial u_i}{\partial t} + \frac{\partial u_j u_i}{\partial x_j} =
    -\frac{\partial p}{\partial x_i} + \frac{1}{\text{Re}}
    \frac{\partial^2 u_i}{\partial x_j \partial x_j}.
    \label{eq:NS}
\end{align}

Equation~\ref{eq:NS} is split into an advection-diffusion equation and a
pressure-Poisson equation using the fractional step method. The presence of a
zero-thickness membrane introduces a Dirac-delta discontinuity. We use the sharp-interface
ghost-cell IBM \cite{mittal2008versatile}, which exactly enforces no-slip and
no-penetration boundary conditions, preserving correct flow structure near the body
\cite{mittal2023origin}. GPU-accelerated solver details are in Kumar et
al.~\cite{kumar2026gpu}.

Membrane deformation is governed by:
\begin{equation}
    M \frac{d^2 \mathbf{X}(t)}{dt^2} = \mathbf{F}_\text{hydro} -
    \mathbf{F}_\text{elas} - \zeta\frac{d\mathbf{X}(t)}{dt},
    \label{eq:bigode}
\end{equation}
where $M \in \mathbb{R}^{N \times N}$ is the diagonal nodal mass matrix,
$\mathbf{X}(t) \in \mathbb{R}^{N \times 3}$ are node positions,
$\mathbf{F}_\text{hydro}$ and $\mathbf{F}_\text{elas} \in \mathbb{R}^{N \times 3}$ are
hydrodynamic and elastic forces, and $\zeta$ is the structural damping coefficient.
The complete procedure is detailed in Kumar et al.~\cite{kumar2026flow}.

\subsection{Added-Mass Stabilization}
\label{app:addedmass}

A fundamental challenge arises when $\rho_s/\rho_f \approx 1$, as in biological
locomotion. The added-mass effect becomes the dominant coupling force, and standard
explicit FSI schemes treating it explicitly lead to unconditional numerical instability.
The total fluid force decomposes as:
\begin{equation}
    \mathbf{F}_\text{hydro} = \mathbf{F}_Q + \mathbf{F}_\kappa, \qquad
    \mathbf{F}_\kappa = -\mathbf{M}_a \frac{d\mathbf{V}}{dt},
\end{equation}
where $\mathbf{M}_a \in \mathbb{R}^{3\times 3}$ is the added-mass tensor with entries:
\begin{equation}
    M_{ij} = \rho_f \int_V \nabla\phi_i \cdot \nabla\phi_j \, dv,
\end{equation}
computed from an auxiliary harmonic potential $\phi$ with Neumann boundary conditions.
We apply semi-implicit diagonal stabilization treating only the dominant $M_{ii}$ terms
implicitly:
\begin{equation}
    \left(m + M_{ii}\right) \frac{V_i^{n+1} - V_i^n}{\Delta t} = F_i^n +
    M_{ii} \frac{V_i^n - V_i^{n-1}}{\Delta t}.
    \label{eq:stab}
\end{equation}
This requires no sub-iterations and adds negligible cost while restoring unconditional
stability, essential for generating high-fidelity training data at biologically relevant
density ratios.

\subsection{Dataset Statistics}
\label{app:dataset}

The full training dataset spans 20 direct numerical simulations covering four membrane
thickness values ($h^* \in \{0.01, 0.02, 0.03, 0.04\}$) and five Strouhal numbers
($St \in \{0.30, 0.325, 0.35, 0.45, 0.50\}$), yielding a $4 \times 5$ parameter grid.
Each simulation provides 200 timesteps of coupled fluid-membrane data. The graph
neighborhood radius for all three edge types is $r = 0.04$, consistent with the
sharp-interface IBM stencil width.

\begin{table}[ht]
\centering
\caption{Dataset summary for caudal fin FSI simulations used to train and evaluate
AeTHERON.}
\label{tab:dataset}
\begin{tabular}{lc}
\toprule
Property & Value \\
\midrule
Number of simulations & 20 \\
Membrane thickness ($h^*$) & $\{0.01,\, 0.02,\, 0.03,\, 0.04\}$ \\
Strouhal number ($St$) & $\{0.300,\, 0.325,\, 0.350,\, 0.450,\, 0.500\}$ \\
Membrane nodes ($|\mathcal{V}_m|$) & 1,757 \\
Membrane triangular elements & 3,355 \\
Fluid nodes ($|\mathcal{V}_f|$) & 24,000,000 \\
Timesteps per simulation & 200 \\
Graph neighborhood radius ($r$) & 0.04 \\
Training window & $t = 0$--$149$ (70/30 random split) \\
Extrapolation window & $t = 150$--$200$ (held out) \\
Flow / membrane loss weight & 0.6 / 0.4 \\
Total dataset size & 906 GB \\ 
\bottomrule
\end{tabular}
\end{table}

\subsection{Model Hyperparameters}
\label{app:hyperparams}

AeTHERON uses separate encoder networks for the fluid and membrane domains before
lifting both into a shared latent space of dimension $d_h = 32$. The processor comprises
$L = 10$ heterogeneous message-passing layers with cross-domain sparse attention of
dimension $d_A = 32$ and sinusoidal time-conditioning
(Equation~\ref{eq:timeconditioning}). The decoder maps the 32-dimensional latent fluid
representation back to the 4-dimensional physical velocity field.

\begin{table}[ht]
\centering
\caption{AeTHERON model architecture and training hyperparameters.}
\label{tab:hyperparams}
\begin{tabular}{lc}
\toprule
Hyperparameter & Value \\
\midrule
\multicolumn{2}{l}{\textit{Architecture}} \\
\midrule
Latent dimension ($d_h$) & 32 \\
Attention dimension ($d_A$) & 32 \\
Time embedding dimension ($d_t$) & 16 \\
Processor layers ($L$) & 10 \\
Fluid encoder: input node features & 4 \\
Membrane encoder: input node features & 10 \\
Fluid kernel width ($k_w$) & 4 \\
Membrane kernel width ($k_w$) & 8 \\
Fluid edge features & 14 \\
Membrane edge features & 7 \\
Total parameters & $\approx$0.6M \\
\midrule
\multicolumn{2}{l}{\textit{Training}} \\
\midrule
Optimizer & Adam \\
Learning rate & $2 \times 10^{-3}$ \\
LR scheduler & Step decay ($\times 0.5$ every 100 epochs) \\
Training epochs & 200 \\
Batch size & 2 \\
Flow loss weight & 0.6 \\
Membrane loss weight & 0.4 \\
Freeze threshold & $10^{-3}$ \\
Training hardware & NVIDIA A100 / L40s, Rockfish (JHU ARCH) \\
Training time & 144 hours \\ 
\bottomrule
\end{tabular}
\end{table}

\subsection{Summary of Flow Results}

\begin{figure}[ht]
    \centering
    \includegraphics[width=0.9\textwidth]{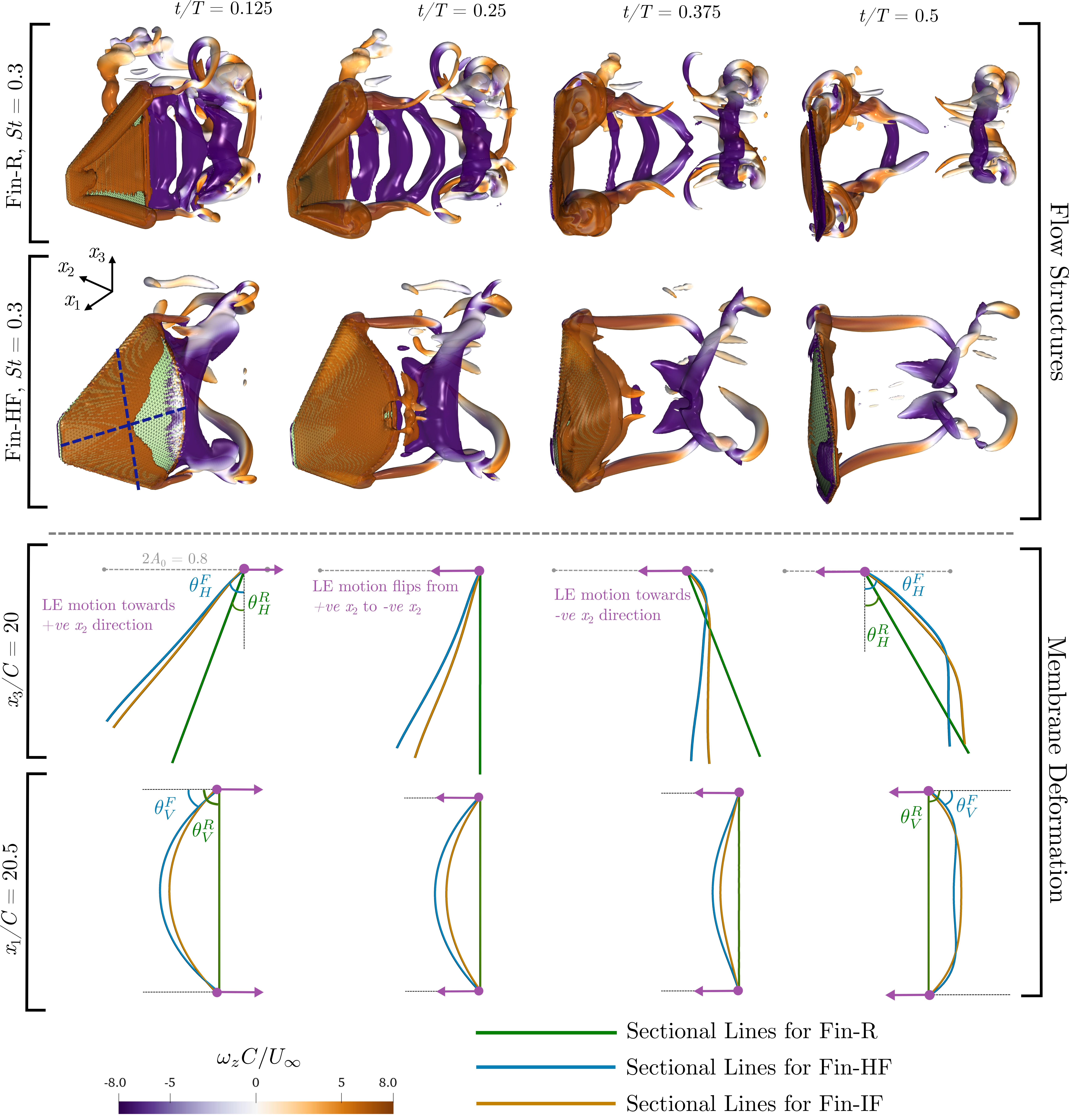}
    \caption{Isosurface of $Q$ at $Q=5$ colored by $\omega_z C/U_\infty$ for three
    representative cases: highly flexible (Fin-HF, $h^* = 0.02$), intermediately
    flexible (Fin-IF, $h^* = 0.04$), and rigid (Fin-R). Top and bottom rows show two
    viewing angles. The flapping cycle progresses left to right, illustrating
    leading-edge vortex formation, mid-stroke shedding, and wake evolution.}
    \label{fig:QIsoVort}
\end{figure}

\begin{figure}[ht]
    \centering
    \subfloat[]{\includegraphics[width=0.3\textwidth]{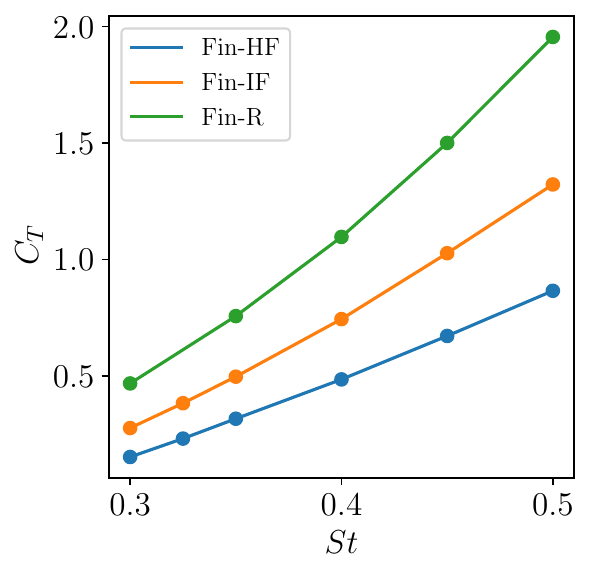}}
    \subfloat[]{\includegraphics[width=0.3\textwidth]{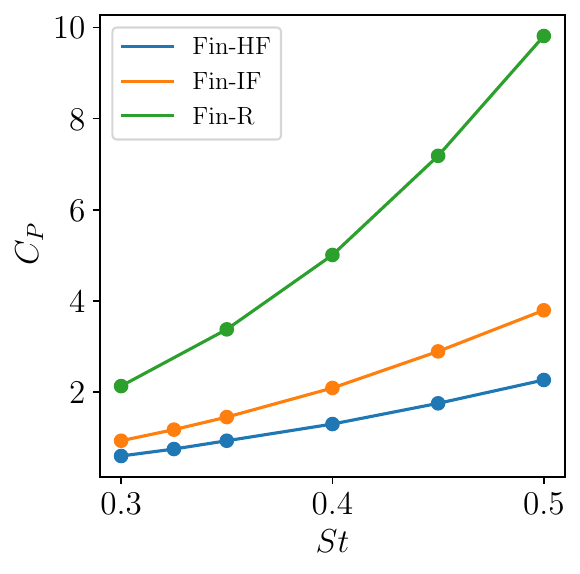}}
    \subfloat[]{\includegraphics[width=0.32\textwidth]{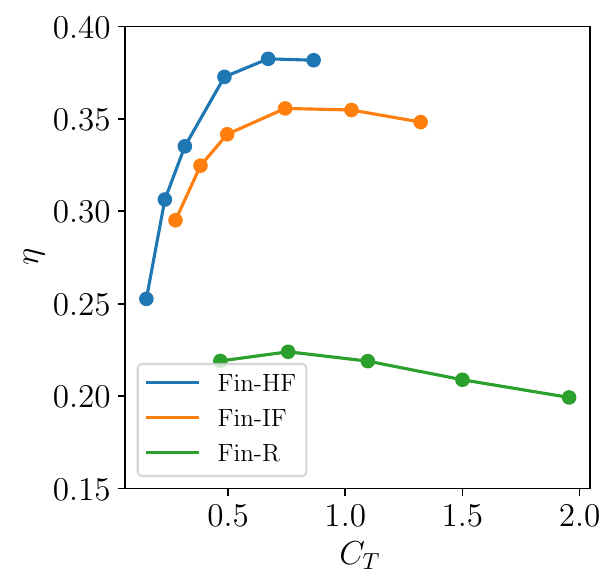}}
    \caption{Hydrodynamic performance across fin flexibility and Strouhal number.
    (a) Thrust coefficient $C_T$ vs.\ $St$, (b) Power coefficient $C_P$ vs.\ $St$,
    (c) Propulsive efficiency $\eta$ vs.\ $C_T$ for Fin-HF ($h^* = 0.02$), Fin-IF
    ($h^* = 0.04$), and Fin-R (rigid). Results shown for $St \in [0.3, 0.5]$. The full
    AeTHERON training dataset spans a broader parameter space; these three configurations
    illustrate the physical diversity captured by the simulator.}
    \label{fig:hydroDataCompiled}
\end{figure}

To provide physical context for the training data, we summarize key flow features and
hydrodynamic performance trends across three representative fin configurations: highly
flexible (Fin-HF, $h^* = 0.02$), intermediately flexible (Fin-IF, $h^* = 0.04$), and
rigid (Fin-R). All results are for $St \in [0.3, 0.5]$, corresponding to efficient
biological locomotion.

Figure~\ref{fig:QIsoVort} shows $Q$-criterion isosurfaces at $Q = 5$, colored by
spanwise vorticity $\omega_z C/U_\infty$, at four flapping phases. All configurations
exhibit the canonical leading-edge vortex (LEV) formation and shedding sequence. Fin-HF
exhibits a more coherent and attached LEV with delayed shedding; Fin-R produces earlier
and more energetic vortex roll-up; Fin-IF shows characteristics of both. These
qualitative differences in vortex dynamics represent the richness AeTHERON must learn
to reproduce.

Figure~\ref{fig:hydroDataCompiled} quantifies performance. Thrust $C_T$ increases
monotonically with $St$ for all configurations, with Fin-R generating the highest raw
thrust. However, $C_P$ tells a different story: Fin-R requires dramatically higher power,
reaching $C_P \approx 10$ at $St = 0.5$ — more than $4\times$ the cost of Fin-HF.
Propulsive efficiency $\eta$ plotted against $C_T$ reveals the decisive advantage of
flexibility: Fin-HF achieves peak $\eta \approx 0.38$--$0.39$ across a broad thrust
range, while Fin-R saturates at $\eta \approx 0.22$--$0.23$. This hierarchy motivates
the choice of $h^*$ and $St$ as the primary parametric axes of the training dataset.

\end{document}